\documentclass[conference]{IEEEtran}
\usepackage{cite}
\usepackage{pgfplots}
\usepackage{amsmath,amssymb,amsfonts}
\usepackage{algorithmic}
\usepackage{graphicx}
\usepackage[braket, qm]{qcircuit}
\usepackage{textcomp}
\usepackage{xcolor}
\usepackage{braket}
\usepackage{physics}
\usepackage{chemfig}
\usepackage{chemformula}
\usepackage{caption}
\usepackage{subcaption}
\usepackage[export]{adjustbox}
\usepackage{float}
\usepackage{subfloat}

\graphicspath{{images/}}

\usepackage{etoolbox}
\makeatletter
\patchcmd{\@makecaption}
  {\scshape}
  {}
  {}
  {}
\makeatletter
\patchcmd{\@makecaption}
  {\\}
  {.\ }
  {}
  {}
\makeatother

\usepackage{hyperref}
\hypersetup{
    colorlinks=true,
    linkcolor=blue,
    filecolor=magenta,      
    urlcolor=cyan
    }

\def\BibTeX{{\rm B\kern-.05em{\sc i\kern-.025em b}\kern-.08em
    T\kern-.1667em\lower.7ex\hbox{E}\kern-.125emX}}

\DeclareUnicodeCharacter{2212}{−}
\usepgfplotslibrary{groupplots,dateplot}
\pgfplotsset{compat=newest}
\tikzset{>=latex}
\usetikzlibrary{
    external,
    patterns,
    arrows,
    positioning,
    chains,
    decorations.pathmorphing,
    decorations.markings,
    decorations.pathreplacing,
    shapes,
    shadows.blur,
    shapes.symbols,
    calligraphy,
    fadings,
    calc,
}
\title{Resource-Efficient Quantum Circuits for Molecular Simulations: A Case Study of Umbrella Inversion in Ammonia}

\author{\IEEEauthorblockN{M. R. Nirmal}
\IEEEauthorblockA{\textit{TCS Research} \\
\textit{Tata Consultancy Services Limited}\\
Bangalore, India \\
mr.nirmal@tcs.com}
\and
\IEEEauthorblockN{Sharma S. R. K. C. Yamijala}
\IEEEauthorblockA{\textit{Department of Chemistry} \\
\textit{Indian Institute of Technology Madras}\\
Chennai, India \\
yamijala@iitm.ac.in}
\and
\IEEEauthorblockN{Kalpak Ghosh}
\IEEEauthorblockA{\textit{Department of Chemistry} \\
\textit{Indian Institute of Technology Madras}\\
Chennai, India}
\and
\IEEEauthorblockN{Sumit Kumar}
\IEEEauthorblockA{\textit{Department of Chemistry} \\
\textit{Indian Institute of Technology Madras}\\
Chennai, India}
\and
\IEEEauthorblockN{Manoj Nambiar}
\IEEEauthorblockA{\textit{TCS Research} \\
\textit{Tata Consultancy Services Limited}\\
Mumbai, India \\
m.nambiar@tcs.com}
}

\DeclareCaptionLabelSeparator{periodspace}{.\quad}
\begin{document}

\maketitle

\begin{abstract}
We conducted a thorough evaluation of various state-of-the-art strategies to prepare the ground state wavefunction of a system on a quantum computer, specifically within the framework of variational quantum eigensolver (VQE). Despite the advantages of VQE and its variants, the current quantum computational chemistry calculations often provide inaccurate results for larger molecules, mainly due to the polynomial growth in the depth of quantum circuits and the number of two-qubit gates, such as CNOT gates. To alleviate this problem, we aim to design efficient quantum circuits that would outperform the existing ones on the current noisy quantum devices. In this study, we designed a novel quantum circuit that reduces the required circuit depth and number of two-qubit entangling gates by about 60\%, while retaining the accuracy of the ground state energies close to the chemical accuracy. Moreover, even in the presence of device noise, these novel shallower circuits yielded substantially low error rates than the existing approaches for predicting the ground state energies of molecules. By considering the umbrella inversion process in ammonia molecule as an example, we demonstrated the advantages of this new approach and estimated the energy barrier for the inversion process.
\end{abstract}


\section{Introduction}
Quantum computing promises computational advantages over its classical counterpart for certain applications. Among the areas expected to benefit from the use of a quantum processor, quantum simulation of molecules comes at the forefront owing to the inherent capability of quantum computers to model systems at the atomic scale \cite{mcardle2020quantum}. However, the numerous limitations of the present-day noisy intermediate-scale quantum (NISQ) computers, such as shorter coherence times, and smaller qubit count, prompt researchers to look for more resource-efficient ways to compute molecular properties using these devices. Since its introduction in 2014 \cite{peruzzo2014variational}, the variational quantum eigensolver (VQE) algorithm has been extensively applied in estimating the ground state energies of molecules on these devices. Although various research groups have demonstrated its efficiency in estimating the ground state energies of small molecules \cite{kandala2017hardware, hempel2018quantum}, VQE is known to provide erroneous energies for large molecules, mainly due to the huge depths of chemically inspired quantum circuits (ansatzes) used in simulating these molecules. 

\begin{figure}[h]
    \centering
    \includegraphics[width=0.9\columnwidth]{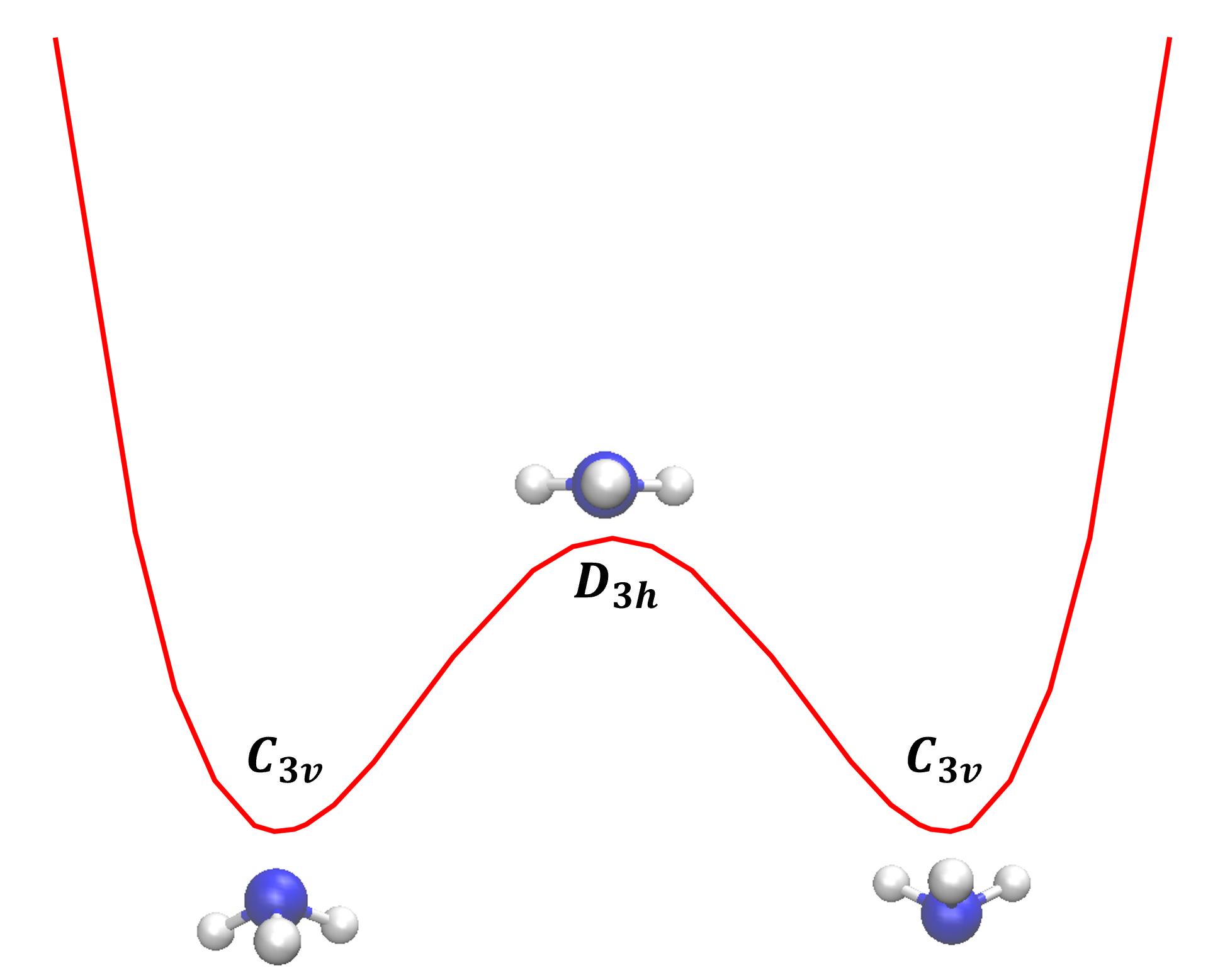}
    \caption{The double minima potential energy curve representing the NH\textsubscript{3} umbrella inversion process.}
    \label{fig: NH3 double minima potential with geometries}
\end{figure}

In this work, we demonstrate how to design shallow ansatzes for VQE algorithm based on Givens rotation gates \cite{anselmetti2021local, arrazola2022universal,ghosh2023deep}. The resulting ansatzes also possess fewer two-qubit entangling gates than the conventional ansatzes. We demonstrate the advantages of the Givens-based ansatzes by simulating the symmetric double minima potential (SDMP) using different types of ansatzes. SDMP is a characteristic of various chemical processes like ammonia (NH\textsubscript{3}) molecule's umbrella inversion process, proton transfer between water molecules, and others \cite{marx2006proton}. Here, we focus on NH\textsubscript{3} and estimate the resources required by each ansatz to accurately compute the energy barrier for its inversion. The inversion process and the associated SDMP are depicted schematically in Fig. \ref{fig: NH3 double minima potential with geometries}. In its equilibrium state, NH\textsubscript{3} has a pyramidal structure with three hydrogens in one plane (H\textsubscript{3} plane) and a nitrogen atom above the H\textsubscript{3} plane, exhibiting a C\textsubscript{3v} point-group symmetry. During the inversion process, the nitrogen atom moves along the three-fold symmetry axis across the H\textsubscript{3} plane through a high energy planar structure, where the molecule exhibits D\textsubscript{3h} symmetry \cite{halpern2007inversion, nirmal2022Evaluating}. This results in an identical C\textsubscript{3v} pyramidal structure, but with the mirror image of the initial structure.

\section{Methodology} \label{methodology}

\subsection{Active Space Selection}\label{ASS}
The electronic Hamiltonian of a molecule, within the Born-Oppenheimer (BO) approximation and using a finite set of molecular orbitals, can be written using the second-quantization formalism as
\begin{equation}
    \hat{\mathcal{H}}_{ele} = \sum_{p,q} h_{pq}a_{p}^{\dagger}a_{q} + \frac{1}{2}\sum_{p,q,r,s} h_{pqrs}a_{p}^{\dagger}a_{q}^{\dagger}a_{r}a_{s}
    \label{eqn: full electronic hamiltonian}
\end{equation}
where, $h_{pq}$ and $h_{pqrs}$ are the one- and two-electron integrals, respectively, computed efficiently on a classical computer. $a_{p}^{\dagger}$ and $a_{p}$ are the canonical fermionic creation and annihilation operators for the orbital $p$, respectively. To simulate chemical systems on a quantum computer, the fermionic operators in \ref{eqn: full electronic hamiltonian} have to be mapped to Pauli operators acting on qubits. Among the existing fermion-to-qubit mapping techniques, we used the Jordan-Wigner (JW) mapping without additional tapering of qubits \cite{bravyi2017tapering} for our simulations. This encoding yields the same number of qubits as there are spin-orbitals in the system. Interested readers are directed to Ref. \cite{seeley2012bravyi} for a detailed review of other encoding schemes, such as Partiy, Bravyi-Kitaev, and other mapping schemes.


In general, while using the JW mapping, owing to the limitations of the NISQ devices, it is not feasible to map all the spin-orbitals of a system onto the qubits. A widely adapted approach in this regard is to choose a few orbitals (and electrons) that contribute significantly to the electron correlation, forming the so-called active space, while taking into account the effect of the remaining orbitals (and electrons) in a mean-field manner \cite{sun2016quantum,stein2016automated}. Traditionally, one chooses the highest occupied molecular orbitals (HOMO) and lowest unoccupied molecular orbitals (LUMO) lying near the Fermi-energy level (HOMO-n to LUMO+m) as the active space. For studying the inversion in NH\textsubscript{3} molecule, we model the system in the minimal STO-6G basis resulting in a total of 10 electrons and 8 molecular orbitals. After running a series of classical simulations, we formed the complete active space (CAS) consisting of two electrons and two orbitals (2e, 2o), by selecting one occupied and one unoccupied orbitals that capture the dominant electronic excitations.


\subsection{Ground State Preparation}\label{GSP}
Gaining a true quantum advantage in simulating molecules using the VQE algorithm involves multiple challenges\cite{tilly2022variational}, such as the need for (a) exponentially large number of measurements for achieving chemical accuracy \cite{gonthier2022measurements}, (b) novel optimization processes to overcome the barren-plateau problem, (c) quantum noise and error mitigation schemes \cite{endo2018practical,cai2022quantum}, and (d) designing a highly expressive and trainable ansatz amenable to the near-term machines. Here, we focus on the challenges associated with the ansatz design. In general, the ansatz has two important attributes, namely, expressibility, and trainability. Here, expressibility refers to the size of the subspace (of Hilbert space) spanned by the ansatz, and trainability refers to the ability in finding the optimal set of parameters of the ansatz that minimizes a given cost function in a tractable time\cite{cerezo2021cost}. Accordingly, the higher the size of subspace, the more expressive the ansatz is\cite{sim2019expressibility,nakaji2021expressibility}, but its trainability could become difficult. Apart from the above, how the circuit depth and the number of two-qubit gates scale with the system size will have a profound impact on the ground state energy predicted using the VQE algorithm on NISQ devices (since they dictate the robustness of the method under noisy conditions). In the remainder of the section, we discuss some of the state-of-the-art approaches to ansatz construction.   


\subsubsection{RyRz Ansatz}\label{RyRz}
\begin{figure}[h]
    \centering
    \includegraphics[width=\columnwidth]{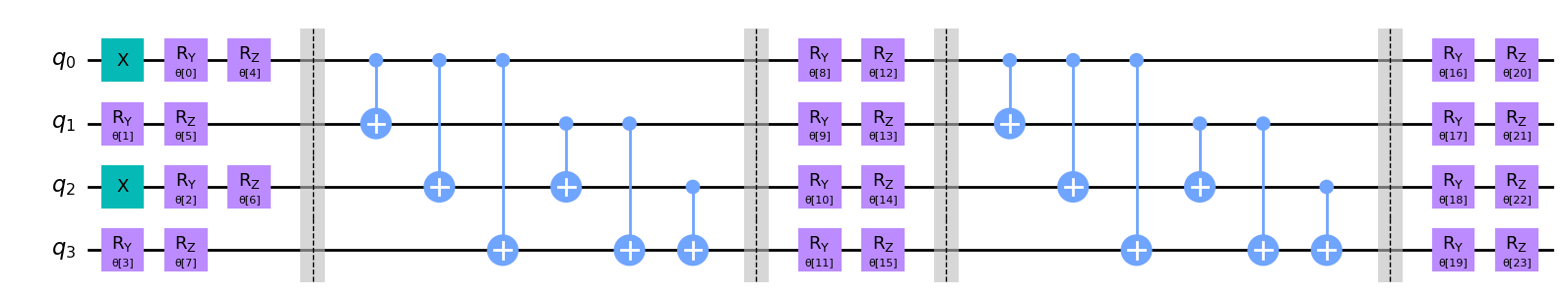}
    \caption{A schematic of the RyRz Ansatz, where the rotation and entangling blocks are repeated twice.}
    \label{fig:RyRz Schematic 2 Reps}    
\end{figure}

This ansatz belongs to the class of hardware efficient (HE) ansatzes that are tailored to use the native gates of the quantum hardware on which the experiment is conducted \cite{kandala2017hardware}. A schematic of the RyRz ansatz with two layers of rotation gates applied on each qubit and two layers of CNOT gates entangling all qubits is shown in Fig. \ref{fig:RyRz Schematic 2 Reps}. Similar circuits are used as parameterized ansatzes in various quantum computing applications, such as machine learning. 

As the name implies, an RyRz ansatz consists of layers of parameterized single-qubit Ry and Rz rotation gates, interleaved with layers of two-qubit CNOT gates. Here, the Ry and Rz gates are given by:
\begin{equation}
    R_{y}(\theta) = e^{-i\theta Y/2} = \begin{pmatrix} \cos{\theta/2} & -\sin{\theta/2} \\    
    \sin{\theta/2} & \cos{\theta/2} \\
    \end{pmatrix}
    \label{eqn:Ry matrix}
\end{equation}
\begin{equation}
    R_{z}(\theta) = e^{-i\theta Z/2} = \begin{pmatrix} e^{-i\theta/2} & 0 \\    
    0 & e^{i\theta/2} \\
    \end{pmatrix}
    \label{eqn:Rz matrix}
\end{equation}
Since the Rz gate applies a complex phase of $e^{\pm i \theta/2}$ to the qubit states, it produces complex-valued wavefunctions. Therefore, in principle, if the interest is to obtain a trail wavefunction with exclusively real amplitudes (as is the case for most molecular ground states), including only the Ry gates in the rotation blocks is sufficient. However, the convergence of such circuits needs to be further analyzed (since more repetitions of the blocks might be required, which increases the circuit depth).


As a general parameterized quantum circuit, the output state of an RyRz ansatz would consist of all the possible $2^n$ computational basis states of $n$ qubits for any arbitrary initialization of the parameters. However, not all these basis states preserve the particle number. Moreover, since the number of electrons in a molecule will be preserved during the measurement, for representing a molecular wavefunction using this ansatz, we need to introduce a penalty term in the cost function to penalize the states with unphysical occupation. This is achieved by constructing a particle number operator $P$
\begin{equation}
    P = \sum_{i} a_i^{\dagger}a_i
    \label{eqn:PN operator}
\end{equation}
where index $i$ runs over all the spin-orbitals and mapping it to qubit operators using JW encoding to yield
\begin{equation}
    \hat{\mathcal{H}}_{pn} = \sum_j h_{j}P_{j} = \sum_j h_{j}\prod_i\sigma_i^j.
    \label{eqn:PN qubit operator}
\end{equation}
The final cost function to be minimized to estimate the ground state of the system becomes
\begin{equation}
    E(\theta) = \bra{\psi(\theta)}\hat{\mathcal{H}}_{elec}\ket{\psi(\theta)} + \mu\bra{\psi(\theta)}\hat{\mathcal{H}}_{pn}\ket{\psi(\theta)}
    \label{eqn:Cost Function}
\end{equation}
where $\mu$ is a parameter, which is large enough to increase the energy of the qubit states that do not preserve the particle number.

\subsubsection{SwapRz Ansatz}\label{SwapRz}
It is one of the HE ansatzes that preserves the particle number\cite{gard2020efficient}. It uses parameterized two-qubit entangling gates of the form given below to entangle all qubits.
\begin{equation}
    R_{XX}(\theta)*R_{YY}(\theta) = \begin{pmatrix}
    1 & 0 & 0 & 0\\
    0 & \cos{\theta/2} & -i\sin{\theta/2} & 0\\
    0 & i\sin{\theta/2} & \cos{\theta/2} & 0\\
    0 & 0 & 0 & 1\\
    \end{pmatrix}
    \label{eqn:Rxx-Ryy matrix}
\end{equation}
where $R_{XX}(\theta) = e^{-i\theta X\otimes X /2}$ and $R_{YY}(\theta) = e^{-i\theta Y\otimes Y /2}$.

Unlike the CNOT gates, the combination of $R_{XX}$ and $R_{YY}$ gates allows the degree of entanglement between two qubits to be controlled by tuning the parameter $\theta$. Also, from the matrix form (see Eq. \ref{eqn:Rxx-Ryy matrix}), it is apparent that this two-qubit gate doesn't modify the $\ket{00}$ and $\ket{11}$ states (columns 1 and 4), and it produces a linear combination only between the $\ket{01}$ and $\ket{10}$ states, which contain the same number of electrons (note that we are in JW representation). Thus, this two-qubit gate conserves the particle number. In the quantum circuit, a layer of such two-qubit gates is sandwiched between two layers of $R_z$ gates that are applied on each qubit. Therefore, similar to the RyRz ansatz, due to the presence of Rz gates, the SwapRz circuit also yields complex-valued states. Moreover, since the SwapRz ansatz is applied after initializing the qubits in the Hartree-Fock (HF) state, an $n$-qubit SwapRz circuit with $m$ qubits excited to $\ket{1}$ state (by applying Pauli-X gates) will give an output consisting of ${n \choose m}$ basis states. Although this ansatz preserves the particle number symmetry, since the entangling gates are applied between the spin-up and spin-down qubits, it does not conserve the spin symmetries.

\subsubsection{Unitary Coupled Cluster (UCC) Ansatz}
This ansatz belongs to the class of chemically inspired ansatzes that are typically derived from classical computational chemistry methods. In particular, the UCC family of ansatzes uses a unitary version of the coupled cluster (CC) theory so that the resulting operators can be directly implemented on a quantum computer \cite{anand2022quantum}. This involves applying an exponentiated unitary operator on some reference state $\ket{\psi_{Ref}}$ to generate the excited electronic states
\begin{equation}
    \ket{\psi_{UCC}} = e^{T-T^{\dagger}}\ket{\psi_{Ref}}
    \label{eqn:Psi UCC}
\end{equation}
where $T$ is the excitation operator, which is given by the sum over singles ($T_1$), doubles ($T_2$), $\ldots$ excitation operators, i.e., $ T = \sum_{i=1}^{n}T_i$. These operators $T_1$, $T_2$, $\ldots$ excite one, two, $\ldots$ electrons, respectively, from the occupied orbitals (indexed as $i, j,..$) to unoccupied orbitals (indexed as $a, b,..$) in the reference state, i.e.,  $T_1 = \sum_{i,a}t_i^a a_a^\dagger a_i$, and $T_2 = \sum_{i,j,a,b}t_{ij}^{ab} a_a^\dagger a_b^\dagger a_i a_b$. Truncating $T$ at doubles (i.e., if $T= T_1+T_2$) results in the popular UCCSD ansatz. The UCC ansatz features all the advantages of the CC method with the added advantage of being variational \cite{taube2006new}. 


\subsubsection{Givens Rotation-based Ansatz}
\begin{figure}[h]
    \begin{centering}
    \includegraphics[width=0.35\textwidth]{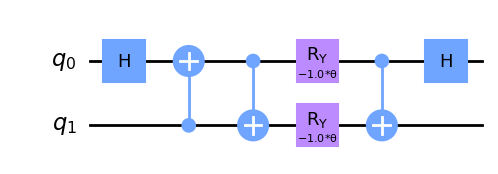}
    \caption{Implementation of the two-qubit hop gate.}
    \label{fig:hop gate}    
    \end{centering}
\end{figure}

\begin{figure}[h]
    \begin{centering}
    \includegraphics[width=0.35\textwidth]{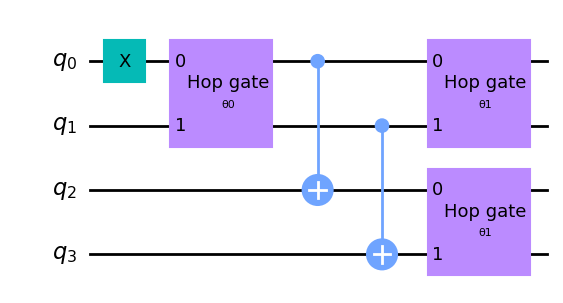}
    \caption{A 4-qubit Givens rotation circuit constructed using hop gates and CNOTs to efficiently prepare the ground state wavefunction of NH\textsubscript{3} molecule in an active space CAS(2e, 2o).}
    \label{fig: Givens_SE_CNOTs}    
    \end{centering}
\end{figure}

Two major bottlenecks associated with the UCCSD ansatz are the quadratic scaling of circuit depth and the fourth-order scaling of CNOT gates with the number of spin-orbitals \cite{tilly2022variational}, making it impractical to run VQE for larger systems on NISQ devices. An alternative approach proposed in the literature utilizes the idea of constructing circuits with particle-conserving unitary gates based on Givens rotation matrices \cite{anselmetti2021local,arrazola2022universal}. This class of circuits consists of single-, double- (and possibly higher-order) excitation gates that perform unitary rotations confined to the subspace of Hilbert space with a fixed number of particles. The basic building block used in our Givens rotation-based ansatz is a two-qubit hop gate that exchanges an electron between two qubits on which it is applied (assuming one of the qubits is excited). The circuit decomposition of the hop gate in terms of Hadamard, Ry, and CNOT gates is shown in Fig. \ref{fig:hop gate} and the matrix representation of the gate operation is given in Eq. \ref{eqn:hop gate matrix}.

\begin{equation}
    h(\theta) = \begin{pmatrix} 
    1 & 0 & 0 & 0\\
    0 & \cos{\theta} & \sin{\theta} & 0\\
    0 & -\sin{\theta} & \cos{\theta} & 0\\
    0 & 0 & 0 & -1\\
    \end{pmatrix}
    \label{eqn:hop gate matrix}
\end{equation}
The operation of the hop gate closely resembles that of the Swap gate (see Eqn. \ref{eqn:Rxx-Ryy matrix}) except for a global phase, and hence conserves the electron number. If we consider CAS (2e,2o) as the active space, then the HF state corresponds to the state where the two electrons are occupied by the lowest energy orbital (between the two orbitals). Now, to recover the full electron correlation energy, we need to add the contributions of one double and two single excitations to the HF state. Referring to Fig. \ref{fig: Givens_SE_CNOTs}, our design of the Givens rotation-based ansatz starts with applying a Pauli-X gate on the spin-up qubit $q_0$, representing the occupation of the lowest energy spin-up orbital. To excite the electron from $q_0$ to $q_1$, we apply a parameterized hop gate between them. A set of two CNOTs following the hop gate entangles the spin-up qubits with the corresponding spin-down ones, effectively creating a doubly excited state. Finally, two hop gates with the same parameter are applied on the pairs ($q_0$,$q_1$), ($q_2$,$q_3$) to create corresponding single excitations.

\section{Results and Discussion}\label{Results}
We employ the ansatzes described in Sec. \ref{GSP} for computing the SDMP representing the inversion process in NH\textsubscript{3}. The software packages used and other computational details are described as follows. The electron integrals $h_{pq}$ and $h_{pqrs}$ defining the electronic Hamiltonian (Eq.\ref{eqn: full electronic hamiltonian}) are computed in the STO-6G basis by running a restricted HF calculation with the PySCF package \cite{sun2018pyscf}. Following this, a complete active space CAS (2e,2o) was chosen as discussed in Sec. \ref{ASS} and mapped the resulting four-spin-orbital reduced Hamiltonian onto qubit operators with the JW mapping, leading to a four-qubit Hamiltonian. The mapping to qubit space and the construction of ansatzes are done with the help of Qiskit software\cite{anis2021qiskit}. 
The accuracy of VQE runs is calculated in terms of absolute errors in energy relative to the CASCI classical method for the chosen active space \cite{helgaker2013molecular}.

\begin{figure}[h]
    \centering
    \includegraphics[width=0.9\columnwidth]{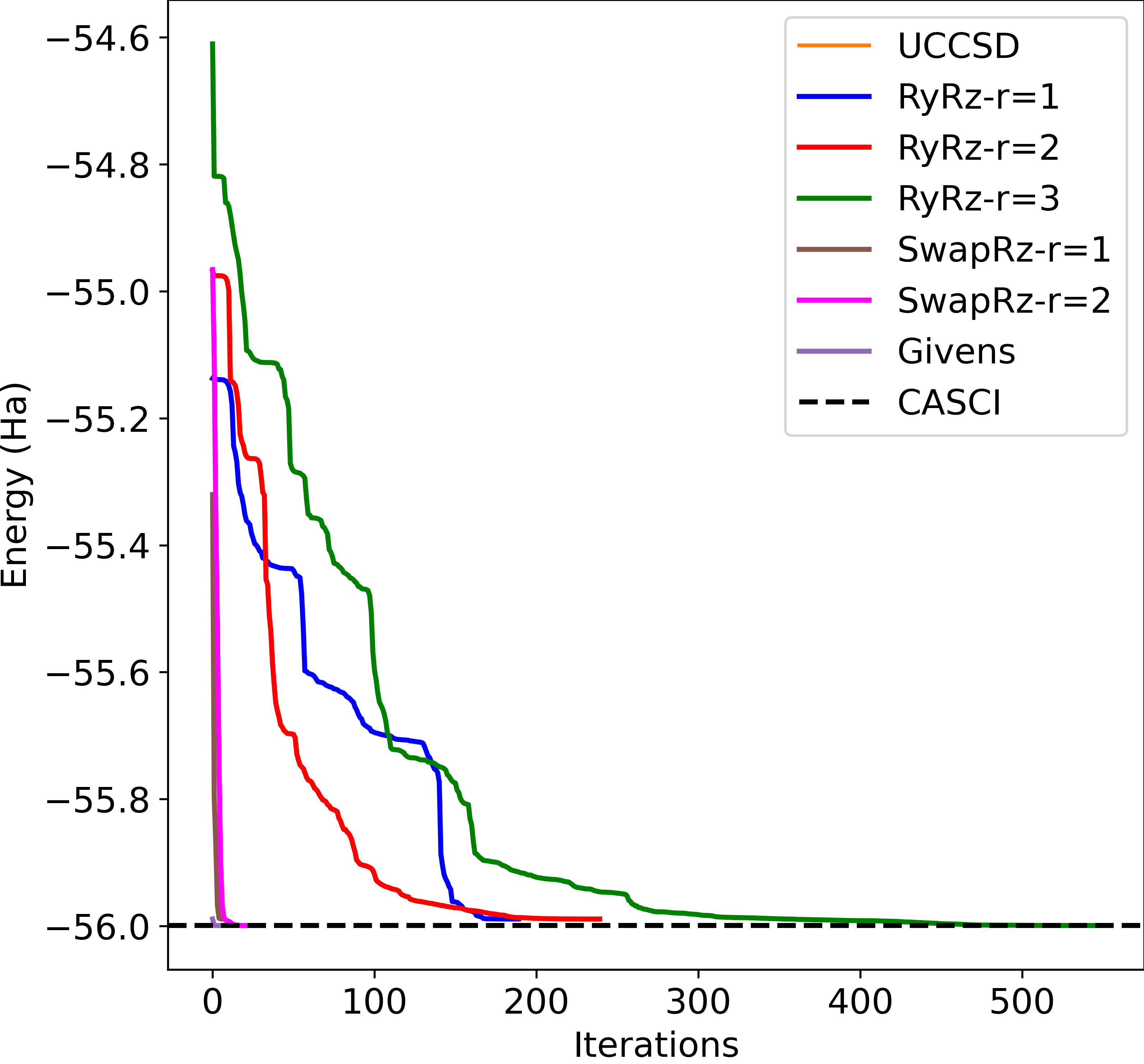}
    \caption{VQE optimization profile showing the convergence of ground state energies with RyRz, SwapRz, UCCSD, and Givens circuits for one geometry of NH\textsubscript{3} and an active space CAS(2e, 2o). RyRz circuits with two and three repetitions (r=1,2) and SwapRz ansatz with one repetition (r=1) gave an absolute error of 10 mHa with respect to the exact (CASCI) energy shown as the dashed black line.}
    \label{fig: Ideal VQE Opt Profile All Ansatzes }    
\end{figure}

The ideal noiseless VQE simulations (statevector simulator of Qiskit) utilized the L-BFGS-B optimizer from the SciPy package \cite{virtanen2020scipy}. For the VQE runs with RyRz ansatz, the penalty parameter $\mu$ in Eq. \ref{eqn:Cost Function} was chosen empirically to be $10^5$. For both UCCSD and Givens rotation circuits, all the parameters were initialized to zero to start from the HF state. However, in the case of the HE ansatzes, initializing to all zeros resulted in an abrupt termination of the optimization without updating the parameters, whereas starting from a set of random values showed successful convergence. With RyRz ansatz, the VQE algorithm converged to the exact ground state energy of the chosen CAS only after three repetitions (r=3 in Fig. \ref{fig: Ideal VQE Opt Profile All Ansatzes }) requiring 32 parameters. This implies, for repetitions less than 3, the RyRz heuristic ansatz is not parameterized sufficiently to recover the correlation energy of $\sim$ 10 mHa.  

On the other hand, convergence to the true ground state was observed for the SwapRz ansatz with 2 repetitions and 14 parameters. The fewer parameters required by the SwapRz circuit compared to RyRz ansatz could be attributed to its particle number-preserving nature as the search space is reduced. In general, we find that if HE ansatzes are used along with larger active spaces (which generally yield greater correlation energies), then a higher number of repetitions of the rotation and entangling layers is required for estimating accurate ground-state energies. These results suggest that the HE ansatzes might become more resource-expensive (or less efficient) for simulating molecules with strong electronic correlations on the NISQ hardware. 

\begin{figure}[t]
    \centering
    \includegraphics[width=0.9\columnwidth]{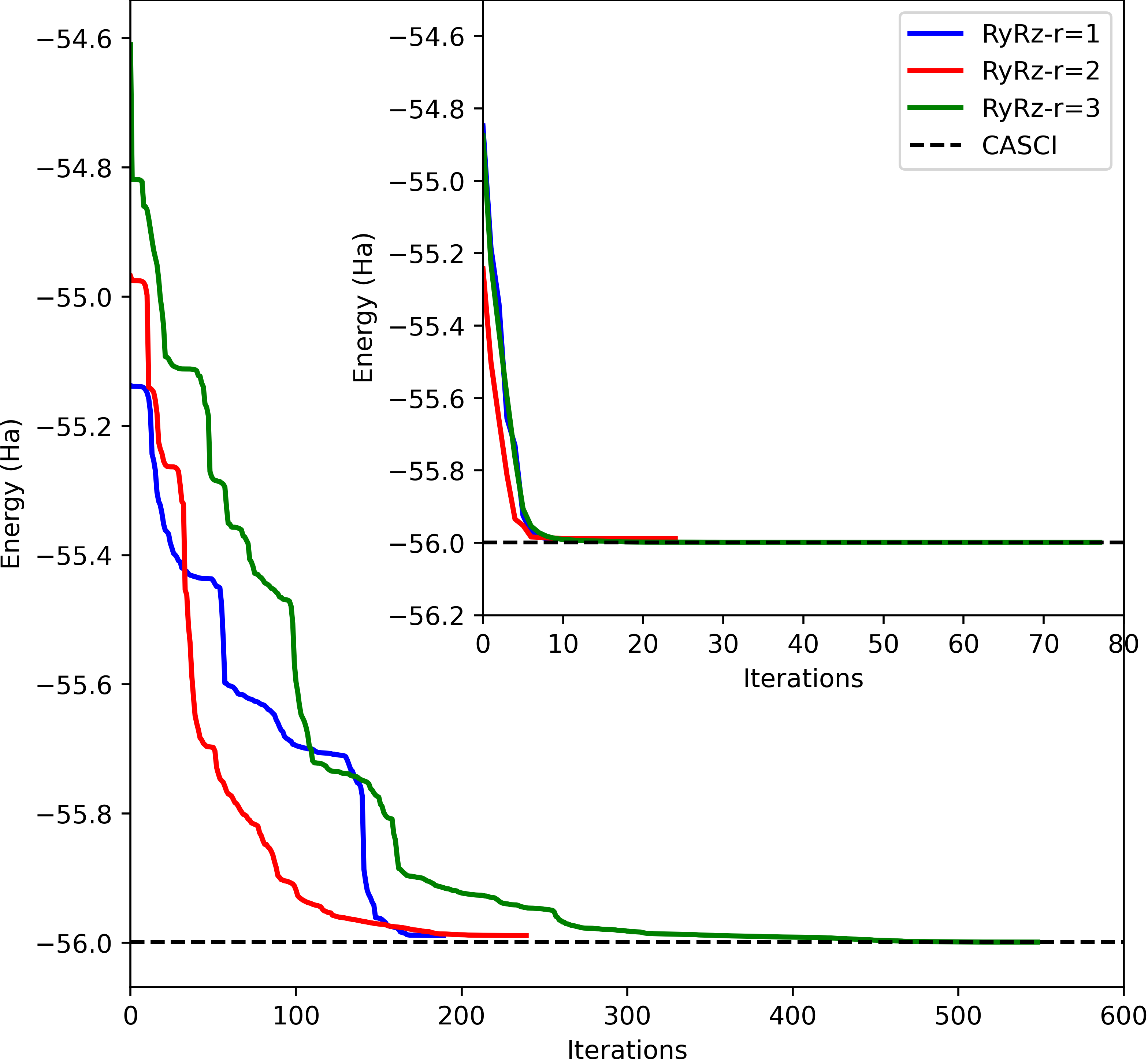}
    \caption{Convergence of VQE optimization using the RyRz ansatz with the penalty term ($\mu=10^5$), and in inset, without the penalty term ($\mu=0$).}
    \label{fig: VQE Opt Prof RyRz with and without penalty}
\end{figure}

As evident in Fig. \ref{fig: Ideal VQE Opt Profile All Ansatzes }, we observed a slower convergence for the RyRz ansatz compared to the SwapRz, UCCSD, and Givens circuits. We find that this slower convergence is due to the presence of the penalty term in the cost function; when the VQE simulations were repeated by setting the penalty term to zero ($\mu=0$), we observed a successful termination with fewer iterations as depicted in Fig. \ref{fig: VQE Opt Prof RyRz with and without penalty}. The faster convergence observed with UCCSD and Givens circuits could be due to the fewer number of parameters in these circuits. 

\begin{figure}[h]
    \centering
    \includegraphics[width=0.9\columnwidth]{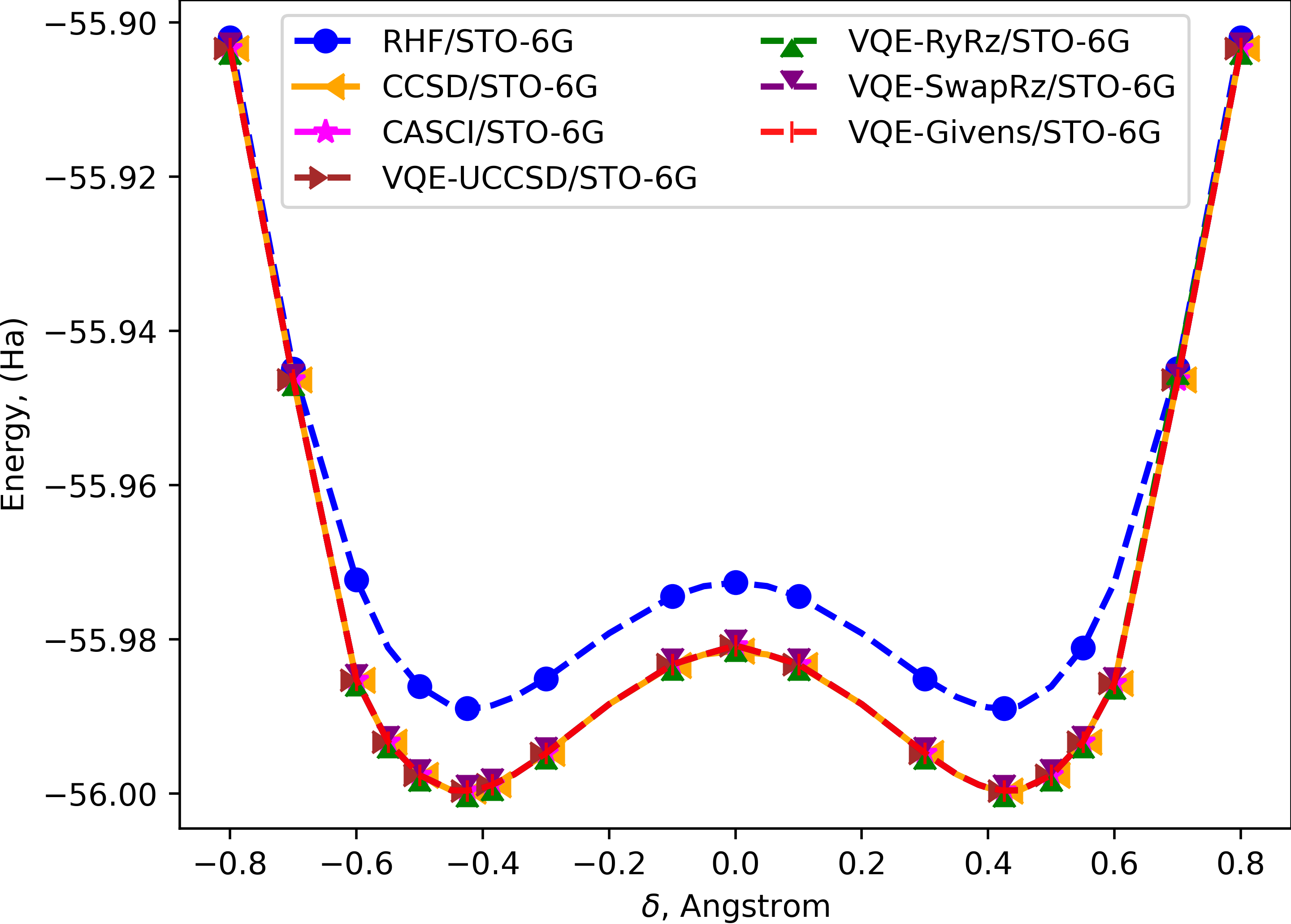}
    \caption{The symmetric double minima potential (SDMP) representing umbrella inversion in ammonia molecule simulated with different classical methods and using VQE with different ansatzes. The $\delta$ coordinate denotes the distance of the nitrogen atom from the plane of hydrogen atoms.}
    \label{fig: NH3 SDMP }    
\end{figure}

After obtaining accurate energy of the NH\textsubscript{3} molecule at its equilibrium geometry with all ansatzes, we proceeded to simulate the SDMP associated with NH\textsubscript{3}'s umbrella inversion. The SDMP shown in Fig. \ref{fig: NH3 SDMP } illustrates the variation in potential energy of NH\textsubscript{3} as the nitrogen (N) atom moves along the symmetry axis across the H\textsubscript{3} plane. For each position of N relative to the H\textsubscript{3} plane, denoted as $\delta$, we performed constrained optimization of the entire molecule at the B3LYP/cc-pVTZ level of theory using the GeomTRIC package \cite{wang2016geometry}. We ran VQE calculations with each ansatz for all these geometries. As expected, we observed two global minima structures for NH\textsubscript{3} at $\delta=\pm0.425 \AA$. The transition state, with D\textsubscript{3h} symmetry, occurs when the N atom becomes coplanar to the H atoms ($\delta=0$). The geometrical parameters and energies of these stationary points are reported in Table \ref{tab: Geometric and Energetics}. From our VQE calculations, we find an energy barrier of 18.807 mHa between the two C\textsubscript{3v} minima, which is consistent with the CCSD and CASCI results obtained on a classical computer. We note that this tiny barrier can facilitate the quantum tunneling of N atom from one minimum to another\cite{halpern2007inversion}.

\begin{table}[h]
    \centering
        \caption{Geometrical parameters and ground state energies of $NH_3$ at the minima and local maximum of SDMP computed with VQE in STO-6G basis. Experimental values are given in the brackets.}
    \begin{tabular}{|c|c|c|}
        \hline
        Property & NH\textsubscript{3} ($C_{3v}$) & NH\textsubscript{3} ($D_{3h}$)\\
        \hline
        $R_{NH}/\AA$ & 1.0190 [1.0124] & 0.9966 \\
        $\angle_{HNH}/\deg$ & 103.83 [106.67] & 120.00\\
        E (Ha) & -55.9995999 & -55.9807928\\
        \hline
    \end{tabular}
   \label{tab: Geometric and Energetics}
\end{table}

\begin{figure}[h]
    \centering
    \includegraphics[width=0.9\columnwidth]{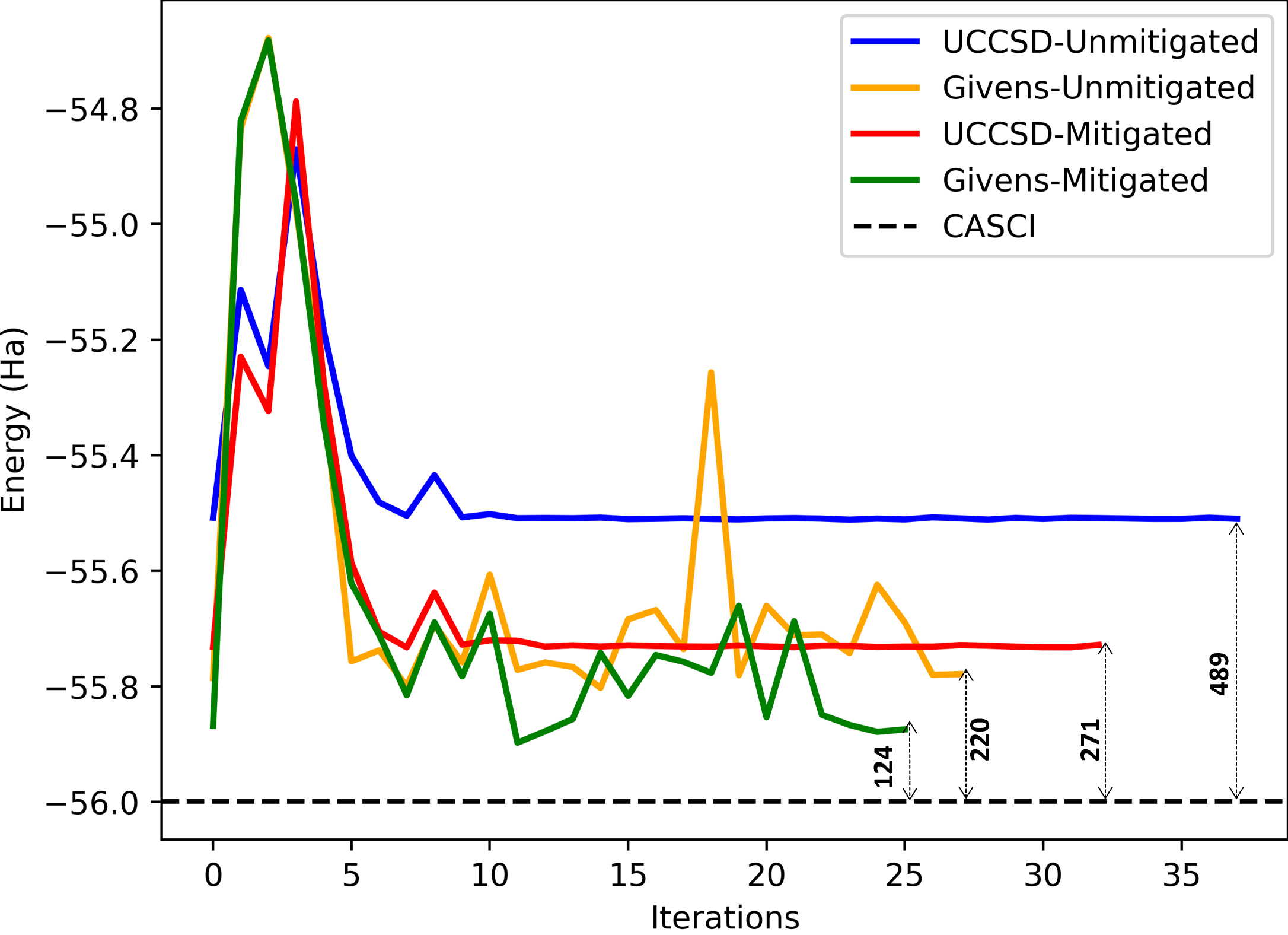}
    \caption{Energy profiles of an NH$_3$ molecule computed using the VQE algorithm with UCCSD and Givens rotation ansatzes in the presence of 127-qubit \textit{ibm\_brisbane} device noise. The energy errors shown are in millihartrees.}
    \label{fig: Noisy VQE }    
\end{figure}

In Table \ref{tab: Quantum Resource estimates}, we reported the quantum resource estimates for each ansatz after transpiling them to the basis gate set of 127-qubit \textit{ibm\_brisbane} device. Clearly, the Givens ansatz provides a significant reduction in circuit depth ($\sim$ $62\%$) and CNOT gate count ($\sim$ $60\%$) compared to the UCCSD circuit while retaining energies within the range of chemical accuracy. To investigate whether this reduction in quantum circuit size would benefit molecular simulations on current quantum computers, we performed VQE simulations in the presence of realistic quantum hardware noise with both circuits. We utilized the \textit{ibmq\_qasm\_simulator} of Qiskit Runtime and configured it with the noise model, coupling map (qubit layout), and basis gate set of \textit{ibm\_brisbane} device. For the VQE simulations with noise, we utilized the COBYLA optimizer, which follows a gradient-free approach and minimizes the errors associated with evaluating numerical gradients of noisy expectation values. For both ansatzes, we initialized the parameters to zeros and used 256k shots to measure the energy at each VQE iteration. We applied the Twirled Readout Error Extinction (TREX) method \cite{van2022model} to mitigate the errors. Although the absolute errors in the energy obtained with both the ansatzes are on the order of hundreds of millihartrees, the error associated with the Givens ansatz is roughly half of the UCCSD ansatz (Table \ref{tab: Quantum Resource estimates} and Fig. \ref{fig: Noisy VQE }), where the improvement is due to the reduced size of the Givens ansatz. Despite the low error observed with the Givens ansatz, considering the large errors ($> 100$ mHa), we find that the present-day quantum computers are not in a position to simulate energy profiles of any important molecular process with an energy barrier of a few milli-Hartrees. As such, more advanced error mitigation strategies need to be developed both at the hardware and algorithmic levels to examine such molecular processes with NISQ devices. 

\begin{table}[h]
    \centering
        \caption{Quantum resource estimates of 4-qubit UCCSD, RyRz, SwapRz, and Givens rotation circuits after transpiling them to a 127-qubit \textit{ibm\_brisbane} device.}
    \begin{tabular}{|c|c|c|c|c|}
        \hline
        Ansatz & Parameters & Depth & CNOTs  & Abs. Error (mHa)\\
        \hline
        UCCSD & 3 & 188 & 43 & 270.63\\
        RyRz(r=3) & 32 & 175 & 42 & 828.07\\
        SwapRz(r=2) & 24 & 323 & 61 & 454.12\\
        Givens & 2 & 71 & 17 & 124.24\\
        \hline
    \end{tabular}

    \label{tab: Quantum Resource estimates}
\end{table}

\section{Conclusion and Future Directions}
In this work, we studied the performance of both hardware efficient (HE) and chemically motivated ansatzes for simulating the symmetric double-well potential, which is a representative of the umbrella inversion process in ammonia molecule. We find that the HE ansatzes require an increasing number of repetitions of the rotation and entangling layers to reach the chemical accuracy, making them as deep as the chemically inspired ansatzes like UCCSD. We presented a novel circuit design with particle-conserving unitary gates derived from the notion of Givens rotation. We showed that this novel circuit design reduces the circuit depth and number of two-qubit entangling gates by almost 60\% while retaining the ground state energies at chemical accuracy. Also, we find lower error rates with this compact circuit compared to the UCCSD ansatz in simulations with realistic device noise.

We plan to extend the design of quantum circuits using Givens rotations to larger active spaces, where we are hoping to obtain similar advantages in terms of quantum resources. As reported recently for interacting spin Hamiltonians \cite{kim2023evidence}, we believe that obtaining very accurate energies on near-term quantum devices might be feasible with the use of sophisticated and advanced error mitigation techniques even with moderately deep circuits. On the other hand, there are methods being developed to reduce the quantum circuit cost based on problem decomposition approaches, such as entanglement forging and circuit cutting \cite{eddins2022doubling}. The scalability of these approaches to tackle larger systems in a tractable time needs to be experimentally validated. 
Another encouraging direction proposed in recent years is the Transcorrelated method that transfers electron correlation from the wavefunction to the Hamiltonian, which yielded accurate energies on the current NISQ devices \cite{sokolov2023orders}.


\bibliographystyle{IEEEtran}
\bibliography{refs}

\end{document}